# Generation of Explicit Knowledge from Empirical Data through Pruning of Trainable Neural Networks

Alexander N. Gorban, Eugeniy M. Mirkes and Victor G. Tsaregorodtsev
Institute of Computational Modeling SD RAS,
Akademgorodok, Krasnoyarsk-36, 660036, Russian Federation
E-mail: gorban@cc.krascience.rssi.ru

**Abstract**

*This paper presents a generalized technology of extraction of explicit knowledge from data. The main ideas are [1)] maximal reduction of network complexity (not only removal of neurons or synapses, but removal all the unnecessary elements and signals and reduction of the complexity of elements), [2)] using of adjustable and flexible pruning process (the pruning sequence shouldn't be predetermined - the user should have a possibility to prune network on his own way in order to achieve a desired network structure for the purpose of extraction of rules of desired type and form), and [3)] extraction of rules not in predetermined but any desired form. Some considerations and notes about network architecture and training process and applicability of currently developed pruning techniques and rule extraction algorithms are discussed. This technology, being developed by us for more than 10 years, allowed us to create dozens of knowledge-based expert systems.*

## 1. Introduction

During training artificial neural network creates some internal rules, but these rules are hidden in the structure of a network and not clear to the user. Moreover, these rules are so difficult for interpretation and understanding. Also discovery of explicit rules for solving of a problem is often more significant than the solution itself. Thus, there is a problem of refining of the hidden knowledge and translating them to natural language.

In this paper we present a generalized three-step technology of extraction of explicit knowledge from empirical data. The eight-year experience of application of this technology to problems in various fields allows us to estimate some propositions and draw come conclusions. The brief description is following.

First, neural network that can solve the problem with desired degree of accuracy has to be trained. Instead of use of mean square error as a estimation function it's better to use modified estimation functions which allow to control accuracy of decision in order to reach more simple resulting structure of the network at the second step of proposed technology. Also, in some cases it would be better to use multilayer neural networks, not only three-layered, in order to achieve more hierarchical set of rules with more simple rules at each level of hierarchy. The successful training of the network creates a hidden complex set of decision rules - the implicit knowledge.

Second, it's necessary to remove superfluous elements and inputs from the network. The pruning should lead to more simple interpretation of hidden rules, therefore it's not enough to remove only neurons or connections and it's necessary to introduce a methodology of complex pruning. So, we present the following set of available pruning operations: removing of inputs, neurons, synapses, biases, uniform simplification of a network (when the maximum number of synapses connected to neuron is decreased over the network). The user can establish the execution order of these operations itself. Then reduction of synaptic weights to values from a finite set of fixed values should be done. All these pruning operations are based on sensitivity analysis, only first-order derivatives are used [1]. Pruning is carried out by consecutive removal of inputs or elements and fails when it's impossible to reach zero of estimation function by retraining. The last step is replacement of sigmoid nonlinear transfer functions by threshold or piecewise-linear functions. Now the hidden implicit knowledge is refined and simplified, so it's possible to understand the meaning and generate explicit knowledge.

The third step is to write down the explicit knowledge in natural language. We propose to carry out such process by consecutive analysis of network structure manually. Because the meaning of input signals is already known, it's possible to substantially name the output signals of the first layer neurons, then second layer neurons and so on. The introduced requirement of network's uniform simplicity is proposed only for simplification of analysis phase, because the less number of input synapses of the neuron the easier to interpret and name its output signal. Also it's possible to extract rules from the network not manually, but automatically in fuzzy form of If-Then form, or in another form. Here we present a complete set of available types of rules that can be extracted from the network. The rule of a certain type may be interpreted in a different ways as a fuzzy, probabilistic or logical statement, but such interpretation involves expert knowledge about problem area and should be done manually by user.

It should be noted that several networks of the same initial architecture often lead to different decision rules (different explicit knowledge). It not seems to be a lack. Created sets of rules begin to compete among themselves, from several sets of rules we can create another one manually.

The rest of the paper explains the proposed ideas more carefully. In Section 2 we describe method of fast training. In Section 3 set of all possible pruning operations is presented. In Section 4 and 5 we discuss existing pruning algorithms and describe our demands and ideas concerning the proposed complex pruning technique. In Section 6 we discuss the possible ways of rule extraction. Section 7 briefly describes the existing applications.

## 2. Fast training of neural networks

It's necessary to note that training of neural network is not a long process. Of course, use of initially proposed back-propagation algorithm [2] leads to hours or days of training time. The main reasons of that and drawbacks of original back-propagation are following:

1. Mean square error (MSE) is used as estimation function. When using MSE it's impossible to stop training when desired accuracy is achieved - training lasts until local minimum is reached. Such situation requires additional stopping criteria (e.g. value of generalization error on test set) and leads to overtraining. Also MSE is the worst estimation for neural classifiers, where rules of interpreting of network output (in order to determine one or another class) are different far from MSE demands.
2. Training in a manner "example by example", where network output for given example is computed and estimated, then gradient vector is computed by back propagation of estimation function derivatives and that gradient vector is added (with given fixed weight - step size) to parameters vector of the network. Then training proceeds to another example. In such a way it isn't possible to use fast optimization algorithms. All the modifications (e.g. adding of momentum term to estimation function) don't significantly speed-up the training and bring the additional difficulties.
3. Using step size without optimization (predefined and fixed for all training time or changed by predetermined or heuristic external law).

But it's possible to speed-up training process up to $10^5$ times [3] in comparison with original back-propagation. The main ideas are following:
1. Training on entire training set (or its sufficiently large subsets) instead of "example by example" training. Total estimation function $H = \sum_{i=1}^{N} H_i$ is optimized where $H_i$ is an estimation of single example and $N$ is a total number of examples in training set. Total gradient (gradient of total estimation function) is a simple sum of gradients of separate examples.
2. Using step optimization methods along the direction of total gradient, for example, parabolic approximation to determine the optimal step sizes.
3. Using fast methods of gradient optimization instead of simple gradient descent. Because it's so difficult to compute second derivatives of total estimation, it's better to use conjugate gradients algorithm or quasi-Newton algorithms (from which BFGS-method with restricted memory requirements is preferable).
4. Using novel estimation functions instead of MSE [4]. For neural predictors it's better to use MSE generalization with predetermined threshold level. For neural classifiers the successful experience of using special estimation functions was summarized in [4].

The main conclusion of this Section - that there is no problems with fast training. It's only necessary to interpret training process as optimization phase and use all the possibilities and well-developed algorithms for fast optimization. The main objectives are requirements of a complicated algorithmic superstructures and going far from basic principle "neural net trains itself". But it seems to be not so expensive demand for substantial speed-up of training, because training of neural net on sets of real data (several hundreds of examples with several dozens of inputs) takes only a few minutes on modern PC.

### 3. Pruning operations

Let's discuss what it's possible to prune from neural network. All the previously published papers didn't introduce a complete set of all possible pruning operations. Here is the list of all operations:
1. Removing input signals from neural network.
2. Removing neurons.
3. Removing synapses.
4. Removing biases. It's important to note that it's better to make a difference between synapses and biases, because bias is a simple constant input but synapse is a weighting line of signal transition and therefore is more difficult to interpret.

That's all the operations that remove signals and units from neural network. From this operations it's possible to construct a higher-level pruning procedures, such as
5. Uniform simplification of the network, when the maximal number of synapses connected to neuron is decreased over the entire network until the given maximal number of synapses is reached. Such operation is carried out by synaptic pruning, but under some rules. This operation is introduced especially for the problem of knowledge extraction, because it's more easy to interpret statement (given by the single neuron) when it's based on a small number (two of three) of input sentences.

Another pruning operations reduce the

complexity of single elements of the network:
6. Replacement of continuous-valued synaptic and bias weights by values from a given finite set of fixed values. The best choice is the set of {-1,1}, but often it's necessary to use more wide sets. The benefit is a chance of possible exclusion of all the weighting coefficients during interpreting the network.
7. Replacement of sigmoid transfer function of neuron by threshold or piecewise-linear or constant function.

It's possible to restrict the scope of every pruning operation from the entire network to layer of neurons or even a single neuron. Here are no strict assumptions about network structure and pruning techniques that realize listed operations. In other words, here's no assumptions how to determine which elements and signals are redundant. Also here's no definition of stopping criteria for pruning process. Such questions are discussed below.

### 4. Pruning algorithms

The existing approaches and ideas of pruning algorithms are following:
1. Adding a penalty term to estimation function. The method was firstly described in [1] (Ishikawa's work on Structural Learning [5] also uses a penalty term for synaptic weights). During training penalized units (synaptic weights) goes to zero and pruning may be done after training in a single step by removing all the synapses and biases whose weights are close to zero. But such approach doesn't provide the flexibility in handling of the pruning process. Also sometimes it's impossible or too difficult to introduce penalties for other elements and signals. Another difficulty arises when combining several penalty terms because of the problem of managing the contribution of each term to resulting estimation function.
2. Removing of neurons that don't contribute to the solution by using some ad hoc rules with rearranging synaptic weights [6,7]. The main idea - the least significantly change the performance (output signals) of the network. More general situation was studied in [8] in order to deal with not only neurons but also every possible subsystem of the network (single synapse, neuron, layer of neurons, substructure of the network etc.).
3. Using so-called sensitivity analysis where sensitivity of estimation function in relation to given modification of the network has to be determined. Then units with least sensitivity can be pruned from the network.

In order to achieve the maximal flexibility of handling and managing the pruning process the last two approaches seems to be the most useful for the problem of knowledge extraction. Second approach doesn't use derivatives of estimation function (and gradients of estimation function on network parameters) but may use derivatives of network units on their inputs or parameters [8]. Third approach may use sensitivities of zero order (which are not based on derivatives of estimation function but use the values of synaptic weights, signals etc. for sensitivity evaluation) or first order sensitivities (which use first derivatives of estimation function on network parameters) or second order sensitivities.

Second order sensitivity analysis was introduced in [9] for synapses (synaptic sensitivity is a multiplication of squared value of synaptic weight and second derivative of estimation function on given weight) and in [10] for neurons (neuron sensitivity is a sum of sensitivities of all input and output synapses of the neuron). Zero order neuron sensitivities were introduces in different forms (some of which are good approximation of the first order sensitivities) in [11-14] and other papers. Zero order synaptic sensitivities are based on the same ideas.

But there exists a paradoxical situation - in common use there's no first order methods of sensitivity analysis. It's caused by the fact that after training (when reaching the local minimum) gradient of MSE estimation function is zero, so it's impossible to use gradient directly. The solution is to average total gradient vector over the several points in parameters space, where averaging points are achieved by small random shift from a local minimum point. Such averaging gives a good approximation of local behavior of estimation function.

When using modified estimation functions which allow to control the desired accuracy of the decision, successful training leads to zero value of total estimation function over the basin that is large enough. Here it's possible to change the accuracy demands for a certain time (by increasing the precision) in order to compute non-zero gradient vector and average it over several training steps.

When averaged total gradient is obtained, first order sensitivity can be computed by multiplication of a gradient component and a given modification of a corresponded parameter. It's a first order approximation of estimation function change caused by given modification.

Duality method [1,8] that is a more general approach to gradient computation than [2] allows to compute gradient of estimation function over all input and intermediate network signals, not only over network parameters. So the sensitivity of network input may be computed as sensitivity of a corresponded signal and sensitivity of neuron - as sensitivity of neuron output. So here is a possibility to use a common approach to prune inputs, neurons, synapses and biases from the network.

The presented approach that use first order sensitivity was introduced 10 years ago [1] and requires much lesser computational time than second-order methods. Moreover averaging over the several points leads first order sensitivities to be a good approximation of sensitivities obtained by second order methods. The experience of expert systems development using first order sensitivity analysis shows the efficiency of this approach [8].

We should point out again that there exists a computationally efficient method of sensitivity analysis that is not empiric (as the most of zero order methods).

In order to replace sigmoid transfer functions of neurons by thresholds or piecewise-linear functions it's possible to use method described in [15] or methods based on second approach listed earlier. So we can do all the pruning operations listed in Section 3.

## 5. Pruning strategies

Let's discuss pruning strategies that may be applicable to the problem of knowledge extraction. The main proposal here is performing pruning by sequential removing or modifying of a single element or signal and retraining network (because of the availability of fast training). This approach allows to perform pruning operations at desired sequence and stop every operation when it's required (not only when it's impossible to achieve zero error by retraining but also when the required number of elements or signals are removed or modified). Also it's possible to adjust a scope for every operation.

The main idea here is to allow user to construct his own pruning strategies by providing him a complete list of pruning operations and specifications. In order to achieve the best results user should select network structure and construct strategy that take into account all the knowledge about problem area. So we have no strict definitions about the pruning strategies because all the real situations are different and there's a lot of possible strategies.

In Section 3 replacement of sigmoid transfer function of neuron by threshold or piecewise-linear function was proposed. Of course, in situations where it's better to deal with other transfer functions (when we have some assumptions about the problem area and/or the type of possible decision) such transfer functions should be placed into the network from the very beginning and corresponding pruning operation should be modified.

## 6. Knowledge extraction

There are many papers concerning knowledge refinement, where existing knowledge first has to be mapped to neural network and after training and pruning reformulated knowledge has to be translated on natural language back. All the techniques discussed here could be applied during training/pruning phase. Also there's many works about insertion of prior knowledge into the network on order to improve performance or decrease training time.

Here we'll focus on phase of knowledge extraction from trained and pruned neural network. One of the possible ways is to extract rules in fuzzy if-then form. A great number of references on that direction were presented in [16]. There are a number of other approaches that don't use fuzzy technique. But there's always exists a difficulty of interpretation of extracted rules in linguistic categories.

All the attempts were made in order to automatize the process of knowledge extraction. But the most difficult is the process of interpretation of extracted knowledge. Such interpretation should be done by user. So return to the beginning (where no automatic extraction was made and all the knowledge extraction from the network was done by user) may be the most appropriate approach. Manual extraction of knowledge may be carried out by consecutive analysis of network structure. Because the meaning of input signals is already known, it is possible to substantially name the output signals of the neurons from the first layer, then from the second layer and so on. The introduced requirement of network's uniform simplicity is proposed only for simplification of the analysis phase, because the less number of input synapses of the neuron, the easier to interpret and name its output signal. In other words, the process of knowledge extraction has being done as a manual building of a structure of symptoms and syndromes (if such medical terms are applicable here). Input signals are first level symptoms, output signals of first layer neurons are first level syndromes second level symptoms and at the same time. Manually it's possible to produce rules of desired form (with or without fuzzification, interpreting the value of a syndrome as a uncertainty factor or real value and so on).

Earlier we pointed out that flexible and adjustable pruning process can help to achieve the knowledge in most appropriate form. Pruning process can be carried out in a manner to produce such resulting network structure from which knowledge in desired form can be extracted. So here we propose one of the possible complete sets of rule types that could be extracted from the network.

A separate neuron generates the corresponded syndrome based on neuron inputs (input symptoms) only. Type and number of rules depend on the type of nonlinear transfer function of the neuron and the type (discrete or continuous) of symptoms. Let's denote output signal (syndrome) as Y and $i$-th value of the syndrome (in the case of discrete syndrome) as $Y_i$. Let's denote input symptoms as $X_1,..,X_n$ where $n$ is a number of input symptoms, and $j$-th value of the $i$-th symptom (in the case of discrete symptom) as $x_{ij}$. $F(X_1,..,X_n)$ is a transfer function. So the following cases are available:
1. If all the symptoms are discrete then syndrome is discrete independently from the type of nonlinear function. For every possible

combination of input symptoms values we can compute corresponded syndrome value and write down the rule in the form of IF ($X_1$=$x_{1j}$ AND $X_2$=$x_{2k}$ AND … AND $X_n$=$x_{nl}$) THEN Y=$Y_i$. The number of rules of given form is specified by the number of possible combinations of symptoms values. If syndrome takes the same value over the several combinations of input values it's possible to combine corresponded rules into one by linking its If-clauses by logical ORs.

All the next situations are related to the case when at least one continuous-valued input symptom is presented.
2. If nonlinear function is smooth then only one functional dependency of the type of Y=F($X_1$,..,$X_n$) can be generated. Of course, here it's possible to use different methods of dividing continuous-valued syndrome on a few discrete linguistic categories, but such work should be done during interpretation phase.
3. If nonlinear function is of threshold type then syndrome is discrete-valued and for every discrete value it's possible to determine restrictions (boundaries) on the weighted sum of input symptoms in the form of IF A<($W_1X_1$+$W_2X_2$+…+$W_nX_n$)<B THEN Y=$Y_i$, where A and B are constants and $W_j$ is a weight of a synapse corresponded to *j*-th symptom. Restrictions may be one-sided and either strict (<) or non-strict (<=). The number of discrete values of the syndrome determines the number of rules. But in the case when for certain combination of values of discrete symptoms any possible change of values of continuous symptoms will not lead to change of syndrome value it's possible to generate rule of type 1 without taking into account all the continuous-valued symptoms.
4. If nonlinear function is piecewise linear then piecewise-constant regions of the function will be described by rules of type 3 and piecewise-linear regions - by rules of type 2.

It's possible to adapt proposed rule set to different situations by interpreting the syndrome values as a certainty factors of probabilities and so on. The main note here is the possibility to determination of sequence of pruning operations that leads to network structure from which it's possible to extracted rules of desired form.

### 7. Applications and results

A great number of expert systems were developed using the proposed algorithms and methods. The problem areas are ecology, psychology and sociology, economy, human health, system's identification and control etc.

One application was presented in [17] where the problem was to make a prediction about the results of US presidential elections (winning of presidential or opposite party) basing on 12 binary questions concerning the pre-election situation. The questions are about economy and social and political situation. We used for training the US presidential elections beginning with the election of 1860 and ending with the election of 1980. While applying the technology of knowledge extraction only five the most significant questions (from initial 12) remained and only two neurons was required to draw the conclusion. Because all the questions had the binary answers ("yes" or "no"), a simple set of rules was created and easily interpreted.

Another result of that investigation was the determination of the most significant questions. Four from 12 initial questions remained the most significant at every run (when difference in network structure or pruning strategy or initial synaptic weights was made). And the fifth minimally required question was different from time to time. The most significant questions were the following:
1. Was there serious competition in the presidential-party primary elections? (The most important question, "yes" - bad for P-

party).
2. Did significant social tension exist during the P-party's previous term? ("yes" - bad for P-party).
3. Was there an average annual growth in the gross national product of more than 2.1% in the last term? ("yes" - good for P-party).
4. Did the P-party president make any substantial political changes during his term? ("yes" - good for P-party).

That result allows us to make a right prediction for all the elections after 1980.

## 8. Conclusion

Here we presented ideas, methods and suggestions concerning the problem of extraction of explicit knowledge from data. The main ideas are:
- first-order sensitivity analysis,
- flexible procedures of pruning of elements that put obstacles in the way of reaching a simple and desired network architecture,
- automatic generation of verbal description of the network.